# Discussion on "Sparse graphs using exchangeable random measures" by François Caron and Emily B. Fox

Mingyuan Zhou

McCombs School of Business

The University of Texas at Austin, Austin, TX 78712, USA

mingyuan.zhou@mccombs.utexas.edu

August 22, 2017

I congratulate Professor Caron and Professor Fox for a well-written paper that establishes a novel statistical network modeling framework, which uses completely random measures to model graphs with various levels of sparsity. Although it has been made clear in the paper that the characteristics of the underlying completely random measure, *e.g.*, the discount parameter $\sigma < 1$ of the generalized gamma process, play a crucial role in determining the sparsity levels of the generated graphs, I should like to call attention to the potentially important role played by the link function $f(x) = 1 - e^{-x}$, in generating unweighted undirected sparse graphs with $z_{ij} \,|\, w_i, w_j \sim \text{Bernoulli}[f(2w_i w_j)]$ for $i \neq j$. With this link function, the contribution of $z_{ij}$ to the negative log-likelihood of the model can be expressed as

$$-z_{ij} \ln[1 - \exp(-2w_i w_j)] + 2(1 - z_{ij}) w_i w_j,$$

which quickly explodes towards $\infty$ as the product $w_i w_j$ approaches zero when $z_{ij} = 1$. Thus the choice of this link function implies an inductive bias towards fitting nonzero edges $z_{ij} = 1$, while not strongly penalizing zero edges $z_{ij} = 0$ even if their corresponding products $w_i w_j$ are large. The same link function, which is referred to as the Bernoulli-Poisson link, has also been used in Zhou (2015), which constructs nonparametric Bayesian network models for overlapping community detection and missing link prediction, allowing the computation to scale linearly with the number of edges, rather than quadratically with the number of nodes. It would be of interest to articulate the role of this specific link function in supporting sparse graphs under the proposed framework.

In addition to controlling for sparse graphs how the number of edges increases as a function of the number of nodes, another topic worth further investigation is how to introduce structured sparsity patterns to the graph adjacency matrices, including modeling dense on-diagonal but sparse off-diagonal blocks, sparse on-diagonal but dense off-diagonal blocks, or a mixture of both. It would be interesting to find out whether the new notion of exchangeability could be maintained while achieving these network modeling goals.